\documentclass[onecolumn,superscriptaddress,aps,pre]{revtex4}

\usepackage{graphicx}
\usepackage{placeins} 
\usepackage{amsmath} 
\usepackage{amsfonts} 
\usepackage{relsize} 
\usepackage{color}
\usepackage{hyperref}

\begin{document}

\title{Spatial maximum entropy modeling from presence/absence tropical forest data}

\author{Matteo Adorisio}
\affiliation{Dipartimento di Fisica e Astronomia ``G. Galilei'', Universit\`{a} di Padova, CNISM and INFN, via Marzolo 8, 35131 Padova, Italy}
\author{Jacopo Grilli}
\affiliation{Dipartimento di Fisica e Astronomia ``G. Galilei'', Universit\`{a} di Padova, CNISM and INFN, via Marzolo 8, 35131 Padova, Italy}
\author{Samir Suweis}
\affiliation{Dipartimento di Fisica e Astronomia ``G. Galilei'', Universit\`{a} di Padova, CNISM and INFN, via Marzolo 8, 35131 Padova, Italy}
\author{Sandro Azaele}
\affiliation{Department of Applied Mathematics, University of Leeds, Leeds, LS2 9JT, UK}
\author{Jayanth R. Banavar}
\affiliation{Department of Physics, University of Maryland, College Park, MD 20742 (USA)}
\author{Amos Maritan}
\affiliation{Dipartimento di Fisica e Astronomia ``G. Galilei'', Universit\`{a} di Padova, CNISM and INFN, via Marzolo 8, 35131 Padova, Italy}

\begin{abstract}
  Understanding the assembly of ecosystems to estimate the number of species at different spatial scales is a challenging problem. Until now, maximum entropy approaches have lacked the important feature of considering space in an explicit manner. We propose a spatially explicit maximum entropy model suitable to describe spatial patterns such as the species area relationship and the endemic area relationship. Starting from the minimal information extracted from presence/absence data, we compare the behavior of two models considering the occurrence or lack thereof of each species and information on spatial correlations. Our approach uses the information at shorter spatial scales to infer the spatial organization at larger ones. We also hypothesize a possible ecological interpretation of the effective interaction we use to characterize spatial clustering.
\end{abstract}

\maketitle

\section{Introduction}

A crucial step in finding a rationale for the overwhelming complexity of an ecosystem is to reliably glean information from  measured quantities. The ever increasing amount of data on real ecosystems are useful to test theories: without these data, there is little to explain and no clear pathway for constructing theories. The availability of detectable and often persistent patterns in nature \cite{condit2002,Volkov2005,Holts2006,Storch2012} has stimulated the scientific community to develop theoretical models that try to explain these regularities.
Along these lines, statistical physics provides powerful theoretical tools and stimulates innovative steps towards the comprehension and unification of the empirical evidence pertaining to the macroecological patterns observed in nature \cite{UNTB,alonso2006merits,blythe2007,Dewar2008,harte2011,rosindell2011,suweis2012a}. 
Indeed, statistical mechanics has been successfully used to develop theoretical frameworks for explaining collective behaviors arising from individual interactions \cite{Chave2002,volkov2003,alonso2004,Azaele2006,volkov2007,Houchmandzadeh2008,bertuzzo2011,suweis2012b}, as in the case of the onset of spontaneous magnetization and diverging correlation lengths in a spin system with local interactions at its critical point. A generalization of the statistical mechanics approach has been proposed to predict the collective behavior of ecological systems \cite{Harte2003}. In fact, ecological systems can be viewed as a set of very large number of interacting entities whose dynamics drives the spontaneous emergence of the system organization.

In this article we propose an inference method belonging to a class of ``inverse problems'' \cite{chayes1984inverse} with the aim of predicting biodiversity patterns over different spatial scales using only the information on local interactions. 
 
In last few decades, extinction rates and biodiversity loss have continued to increase putting at risk several ecosystem benefits: provisioning, supporting, regulating and cultural services. The knowledge of biodiversity spatial patterns is crucial to quantify the extinction rates due to habitat loss.  Most predictions of biodiversity patterns are inferred by applying the species area relationship (SAR) to rates of habitat loss. Recent results showed that the most appropriate method to describe extinction rates are through the endemic area relationship (EAR) \cite{He2011}. Both SAR and EAR are spatial patterns, and indeed the relevance of explicit spatial models in ecology has been very much recognized \cite{legendre1989spatial,kerr2002local}.
In particular the SAR relates the number of distinct species to the sampled area and so describes the increase in species richness with geographic area. On the other hand the EAR expresses the trend between the number of species completely contained (\textit{endemic}) in a given area and the sampled area.

In this work we propose a model to study the spatial structure of a rainforest ecosystem by using the maximum entropy principle. This powerful tool, borrowed from statistical mechanics~\cite{Jaynes2003}, has a wide range of applications~\cite{Burkoff2013,Schneidman2006} especially in the field of ecology~\cite{Harte2009a,Azaele2010}. To our best knowledge, the application of maximum entropy models has never considered space explicitly. Here we propose a method to consider space as an explicit degree of freedom in a maximum entropy model.

One of the major problems in the investigation of an ecosystem’s organization is to understand how the information about the species composition in a portion of the system can be used to characterize species composition at larger areas \textit{(species upscaling)}. Thus a spatially explicit maximum entropy model can be useful to characterize the ecosystem structure and also be adapted to study species upscaling.
Furthermore, the knowledge of species composition is generally restricted to small portions of territory. Upscaling species richness from local data to larger areas is therefore a necessary step to characterize species richness at non-local scales. Maximum entropy models are, by definition, the way to extract the largest amount of possible information from data. In this sense, they represent the natural and possibly most promising way to upscale information on ecosystems composition. This possibility has already been explored with non-spatial maximum entropy models \cite{Harte2009a}, but has not fully exploited. Our spatially explicit approach allows one to infer local spatial interactions that can be used as the building blocks of the spatial distributions of species at any spatial scale.

\section{The Max-Ent principle for the ecological community}

The maximum entropy principle is a useful method to obtain the least biased information from empirical measurements \cite{Jaynes2003,harte2011}.

Let us consider an ecosystem contained in a given area $A$ divided in $N$ adjacent sites of equal area and further
assume that we have knowledge of which species are contained within it. In order to characterize the state of our system, we introduce the binary variable $\sigma^\alpha_i$, which registers the occurrence of each species $\alpha=\{1,\cdots,S\}$ in each site $i=\{1,\cdots,N\}$. 

Each species is thus represented by a vector of binary entries where $\sigma^\alpha_j=0$ and $\sigma^\alpha_j=1$ represent respectively the absence and the presence of the species $\alpha$ in a particular site $j$. All the information about species' occurrences within our ecosystem is contained in the set of vectors $\boldsymbol{\sigma}^\alpha=(\sigma^\alpha_1,\cdots,\sigma^\alpha_N)$, with $\alpha=\{1,\cdots,S\}$.

For the sake of simplicity, let us assume that the occurrence of any species is independent of the other ones. Then the probability of finding a given species $\alpha$ in the configuration $\boldsymbol{\sigma}^\alpha$ can be written as:
$
  \mathbb{P}(\boldsymbol{\sigma}^1,\cdots,\boldsymbol{\sigma}^S  ) = \prod\limits_{\alpha=1}^S p_\alpha(\boldsymbol{\sigma}^\alpha),
$
where $p_\alpha$ is the probability distribution of finding species $\alpha$ in the configuration $\boldsymbol{\sigma}^\alpha$. At  first sight, this assumption may be considered an oversimplification. However, it has been shown that individuals belonging to different species (within a trophic level) can be considered independent to a first approximation \cite{Hoagland1997,volkov2003,Veech2006,Volkov2009,Azaele2010}.

We now use the maximum entropy principle (MaxEnt) \cite{Jaynes2003,banavar2010} in order to characterize $p_\alpha$ and thus the probability to observe the whole system in a given configuration. For simplicity, in what follows, we drop the $\alpha$ index.

From presence-absence data, we can build the \textit{observed configuration} $\hat{\boldsymbol{\sigma}}$.
All the information we know about a given species' occurrence is thus contained in this vector. To build our spatial MaxEnt model we will impose constraints on the average presence of the species in the ecosystem $\hat{M}=\sum_i \hat{\sigma}_i$  and the co-occurrence of the species in neighboring sites $\hat{E}=\sum\limits_{\langle i,j \rangle} \hat{\sigma}_i \hat{\sigma}_j$. In particular, focusing on the co-occurrence in neighboring sites is suggested by the fact that there is evidence that individuals belonging to the same species are spatially clumped \cite{Plotkin2000a,morlon2008}.

We proceed by selecting two indicators that summarize the information contained in $\hat{\boldsymbol{\sigma}}$ and we want to test if such information is sufficient to model the biodiversity patterns of the species in the ecosystem.

The MaxEnt framework \cite{Jaynes2003} maximizes the Shannon's entropy ($\sum _{\boldsymbol{\sigma}} p(\boldsymbol{\sigma}) \ln p(\boldsymbol{\sigma}))$ constrained to match the empirical averages $\hat{M} = \sum_i \hat{\sigma}_i$ and $\hat{E} = \sum\limits_{\langle i,j \rangle} \hat{\sigma}_i \hat{\sigma}_j$ with the ones calculated using  $p(\boldsymbol{\sigma})$. 
It can be shown that in this case $p(\boldsymbol{\sigma})$ takes the form:
\begin{equation}
  p(\boldsymbol{\sigma}| h,J) = \frac{1}{Z(h,J)} \exp \Big( J \sum\limits_{\langle i,j \rangle} \sigma_i \sigma_j +  h \sum_i \sigma_i \Big) = \frac{e^{J E(\boldsymbol{\sigma}) +  h M(\boldsymbol{\sigma})}}{Z(h,J)} ,
\end{equation}
where $ Z(h,J) = \sum\limits_{\boldsymbol{\sigma}} e^{ J E(\boldsymbol{\sigma}) +  h M(\boldsymbol{\sigma}) } $ is a normalization constant called the partition function. In this way, we can characterize the probability distribution for every single species relying only on $M$ and $E$. The parameters $h$ and $J$ must be fitted from the data for each species to satisfy the imposed constraints. The parameter $h$ of each species is taken to be constant over all the sites and can be interpreted as an external factor that favors the presence ($h>0$) or absence ($h<0$) of the species (i.e. analogous to an external magnetic field in the spin model).  $J$ can be interpreted as an effective interaction between neighboring sites that favors the clustering ($J>0$) or the dispersion ($J<0$) of a species among the different sites. A possible interpretation of this effective interaction is a density- or distance-dependent effect of the reproduction rates of tropical tree species due to host-specific predators or pathogens (Janzen-Connell effect) \cite{Janzen1970,Connell1971} with a resulting less clustered distribution.

In the case $J=0$, the model reduces to the case of independent sites that resembles a well studied case known as the random placement model (RPM). The RPM  was introduced in ecology for the first time by Coleman~\cite{Coleman1981} to study spatial patterns in species under different hypotheses on the species abundance distribution. Our framework does not require the knowledge of species abundances, but only the presence-absence information. In this case, the RPM can be built imposing $M$ as the unique constraint and the probability distribution takes the form
\begin{equation}
  \label{eq:probRPM}
  p(\boldsymbol{\sigma}|h) = \frac{1}{Z(h)} \exp \Big( {h \sum_i \sigma_i} \Big) \,
\end{equation}
where $Z(h) = \Big( 1 + e^{h} \Big)^N$. In this case $h$ is fixed by the empirical $\hat{M}$ and can be shown that $h = \ln \left( \frac{\hat{m}}{1-\hat{m}} \right)$ where $\hat{m}=\hat{M}/N$ (see appendix~\ref{app:hMrelation}).

\subsection{Spatial patterns in a presence/absence framework}

In order to characterize the spatial structure of an ecosystem we focus on the species area relationship (SAR) and the endemic area relationship (EAR). The SAR expresses the mean number of species found in a sampled area of an ecosystem. Similarly the EAR can be measured counting the number of species completely contained (endemic) in the sampled area. In our presence/absence framework, we can define the SAR as
\begin{equation}
  SAR(a) = \sum\limits_{\alpha=1}^S \Big \langle  \chi_a(\{\sigma\}) \Big \rangle_{\boldsymbol{g}_\alpha}  = \sum\limits_{\alpha=1}^S SAR_\alpha(a)
  \label{eq:SARinteract}
\end{equation}
where 
\begin{equation*}
  \chi_a(\{\sigma\})  = 
  \begin{cases}
    1,& \text{if} \ \sum_{i \in a} \sigma_i > 0 \\
    0,              & \text{otherwise}
  \end{cases}
\end{equation*}
The EAR is given by
\begin{equation}
  \label{eq:EAR}
  EAR(a) = \sum\limits_{\alpha=1}^S \Big \langle \mathlarger{\delta}_{\sum\limits_{i \in a^c} \sigma_i,0} \Big \rangle_{\boldsymbol{g_\alpha}} = \sum\limits_{\alpha=1}^S EAR_\alpha(a) \,
\end{equation}
where the Kronecker delta imposes the condition of endemicity using the fact that if a species is endemic in an area $a$ it is not present outside of it (i.e. it is not present in $a^c$ the complementary space to $a$).

In both eq.~\eqref{eq:SARinteract} and eq.~\eqref{eq:EAR},  the expression $\langle \cdots \rangle_{\boldsymbol{g}_\alpha} $ stands for the average calculated using the probability distribution $  p(\boldsymbol{\sigma}| \boldsymbol{g_\alpha} )  $ with $ \boldsymbol{g_\alpha} = (h_\alpha,J_\alpha) $

In the case of RPM, we can analytically compute both the SAR and the EAR
\begin{equation}
  SAR_{RPM}(a) = \sum_{\alpha=1}^S SAR_\alpha(a) = S - \sum_{\alpha=1}^S (1-\hat{m}_\alpha)^{|a|} \,
\end{equation}
where $|a|$ is the size of the sampled area (we measure the area in terms of number of sites $|a|$ in the range from $1$ to $A=N$), while 
\begin{equation}
\label{eq:EARrpm}
  EAR_{RPM}(a) = \sum_\alpha \left( 1-\hat{m}_\alpha \right)^{A-|a|}.
\end{equation}
In the RPM case, the EAR and the SAR are related in a simple way~\cite{He2011}
\begin{equation}
  \label{eq:EARfromSAR}
  EAR_{RPM}(a) = S - SAR_{RPM}(A-a) \ .
\end{equation}
We note that, using Eq.~(\ref{eq:EAR}), the EAR can be written as
\begin{equation}
  \label{eq:EAR_withZ}
  EAR_\alpha(a) = \frac{Z_a(h_\alpha,J_\alpha)}{Z_A(h_\alpha,J_\alpha)}
\end{equation}
Here $Z_a(\boldsymbol{g}_\alpha)$ and $Z_A(\boldsymbol{g}_\alpha)$ are the partition functions evaluated respectively in a subconfiguration of area $a$ and on the whole lattice of area $A$. As in the case of the SAR, if $J=0$, we obtain an analytical expression for the EAR. In fact, imposing $J=0$ in Eq.~\eqref{eq:EAR_withZ}, we obtain (see Appendix~\ref{app:EAR_rpm})
\begin{equation}
  \label{eq:EARrpm_withZ}
  EAR_{RPM}^\alpha(a) = \frac{Z_a(h_\alpha,0)}{Z_A(h_\alpha,0)} = \left( \frac{1}{1+e^{h_\alpha}} \right)^{A-|a|}.
\end{equation}
The fact that $h_\alpha$ is fixed by the mean occupation $\hat{m}_\alpha$ allows us to rewrite Eq.~(\ref{eq:EARrpm}) as Eq.~(\ref{eq:EARrpm_withZ}). 

\section{Methods}

\subsection{From spatial data to presence/absence data}
\label{sec:paData}
Presence/absence data were obtained starting from abundances of the Barro Colorado Island (BCI) rainforest (50ha permanent plots). We divided the surveyed area in $N=256$ cells and assigned to each one a variable $\hat{\sigma}^{\alpha}_i$ for $i \in \{1,\cdots,N\}$ and $\alpha \in \{1, \cdots, S\}$ such that $\hat{\sigma}_i^\alpha = 1 $ when species $\alpha$ is present at the cell $i$ and $ \hat{\sigma}_i^\alpha = 0 $ if it is absent.
Applying this procedure we end up with a lattice like configuration, one for each species.

\subsection{Algorithm to find h and J}
In order to find the $h$ and $J$ that reproduce the constraints, we minimize the function 
\begin{equation}
  \label{eq:convexH}
  H(h,J) = \ln Z(h,J) - \hat{M} h - \hat{E} J.
\end{equation}
From an information theory point of view, this function follows from the Kullback-Liebler divergence requiring that the ``distance'' between $p(\boldsymbol{\sigma}|h,J)$ and the unknown real distribution $p_{data}$ must be minimized with the right choice of $h$ and $J$. The fact that the functional is convex~\cite{cover2012} make this task very simple.

\section{Results and discussion}

\begin{figure}[h!]
  \centerline{ \includegraphics[width=.8\linewidth,,height=0.8\linewidth,keepaspectratio]{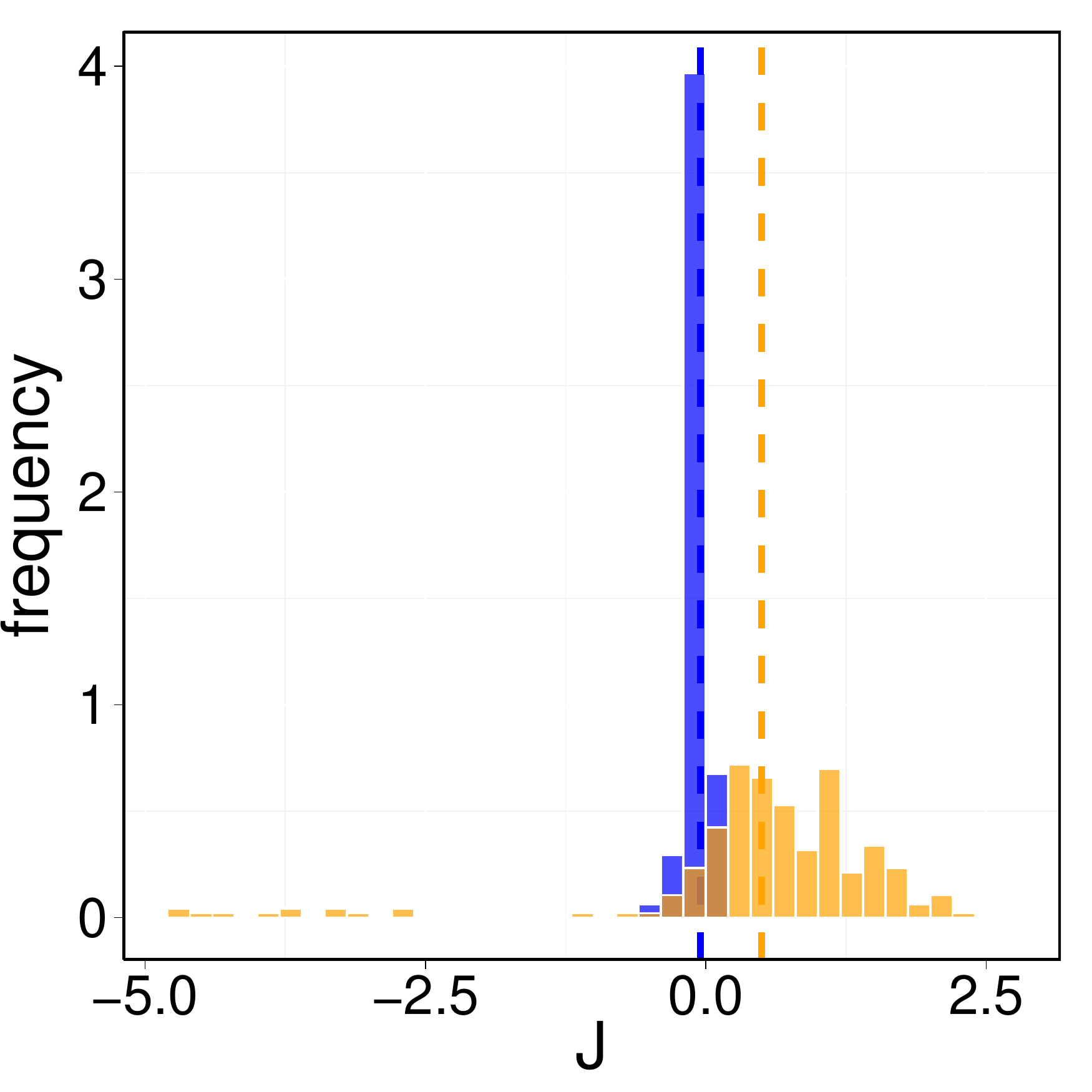} }
  \caption{\small{\textbf{J couplings} Histograms of the coupling strengths $J$ obtained for the interacting model (orange) and for the randomized configuration (blue). As expected, the $J_{rnd}$'s inferred for randomized data have a pronounced peak around zero. The positive mean of the $J$ corresponding to the BCI configuration is a measure of species clustering. The two dashed lines represent the means of the two histograms. The $J$ values in the range [-5,-2.5] are likely related to the Janzen-Connell effect.}}
  \label{fig:Jhisto}
\end{figure}

We applied our approach to the BCI rainforest (see~\ref{sec:paData}) and we examined whether the interacting model is capable of reproducing spatial patterns such as SAR and EAR. To understand the ``degree of randomness'' of the BCI at this spatial scale (fixed by $N=256$), we compare the coupling parameters $J$ and $J_{rnd}$ obtained respectively from the BCI dataset and a random dataset where configurations were generated with the same number of occupied sites but distributed randomly.
As expected the $J_{rnd}$ have almost zero mean and very small variance ($\sigma^2 \approx 10^{-3}$) (Figure~\ref{fig:Jhisto}). In the case of the configurations obtained from real data, despite a few cases where $J_\alpha<0$, it is clear that the $J$ of the interacting model are positive and significantly different from $J_{rnd}$ (see Figure~\ref{fig:Jhisto}). The analysis of $J$ thus suggests that species tend to form clusters. The overall positive mean of the $J_\alpha$ (see Figure~\ref{fig:Jhisto}) produces a clustering of the individuals that has the effect of decreasing the mean number of species with respect to the one of the random placement model. Although our approach does not consider species abundances, the results for the random placement model agree with the conclusions of previous works~\cite{Plotkin2000a} where the authors reported the inadequacy of the random placement model for various rainforest ecosystems. 

\begin{figure}[h!]
  \centering
  \includegraphics[width=0.8\linewidth,height=0.8\linewidth,keepaspectratio]{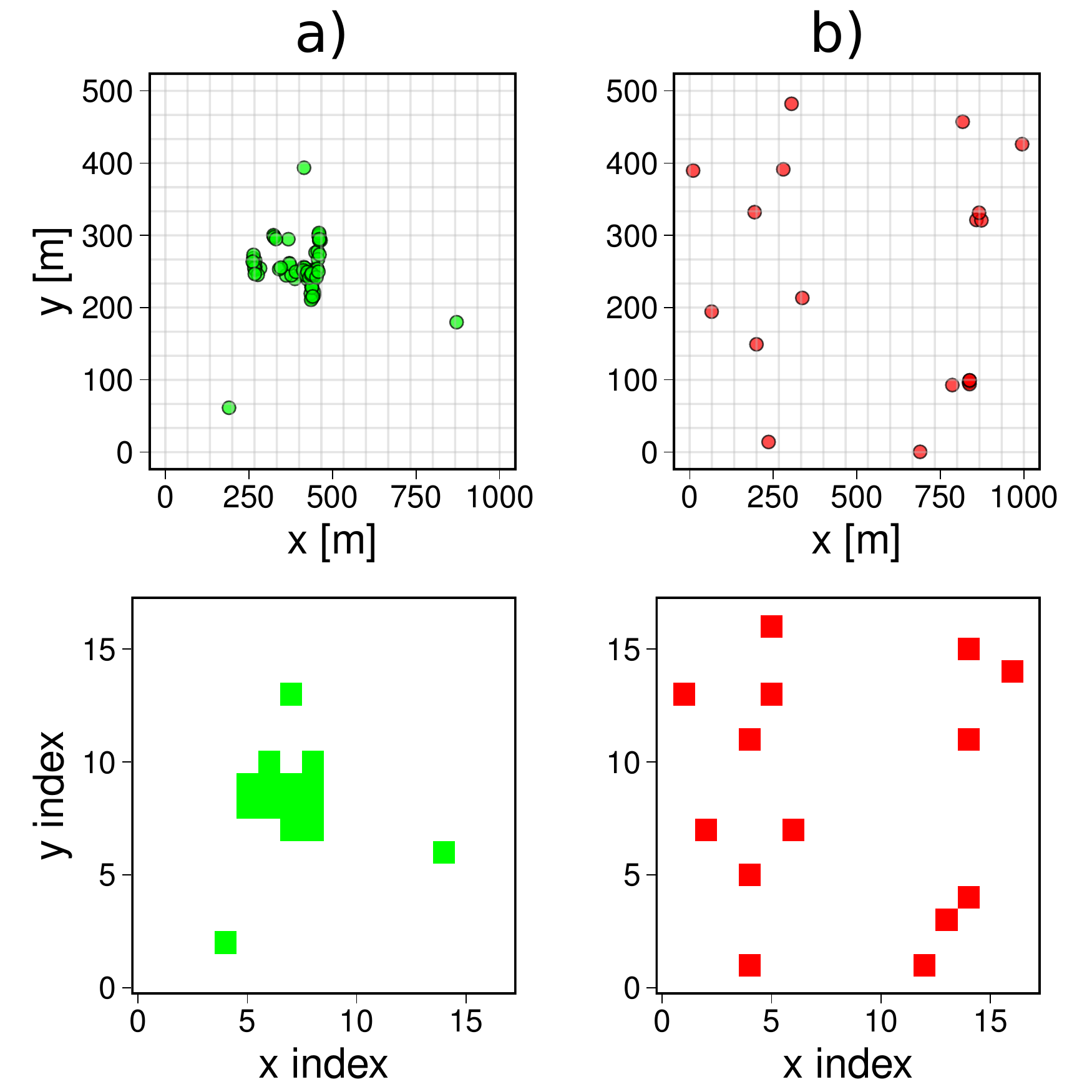}
  \caption{\small{\textbf{Clumping} An example of two species with J values of different signs. Case a) the real spatial distribution (top) of a species with $J_\alpha>0$ and the lattice configuration (bottom) and  b) the same as in case a) for a species with $J_\alpha<0$. In the case a) $\hat{M}_\alpha = 15$ and in the case b) $\hat{M}_\alpha = 14$.}}
  \label{fig:configurations}
\end{figure}

The analysis of the couplings also reveal a possible interpretation for the cases in which the $J$ takes on a negative value. Even if we had analyzed the system at a coarser spatial scale than the one of the single individual, we find that some species, for which the Janzen and Connell effect has been reported, are characterized by a negative $J$ corresponding to a less clustered configuration. Figure~\ref{fig:configurations} shows an example of two species with similar $\hat{M}_\alpha$ but opposite $J_\alpha$ values.\\

\begin{figure}[h!]
  \centerline{ \includegraphics[width=1.2\linewidth,,height=1.2\linewidth,keepaspectratio]{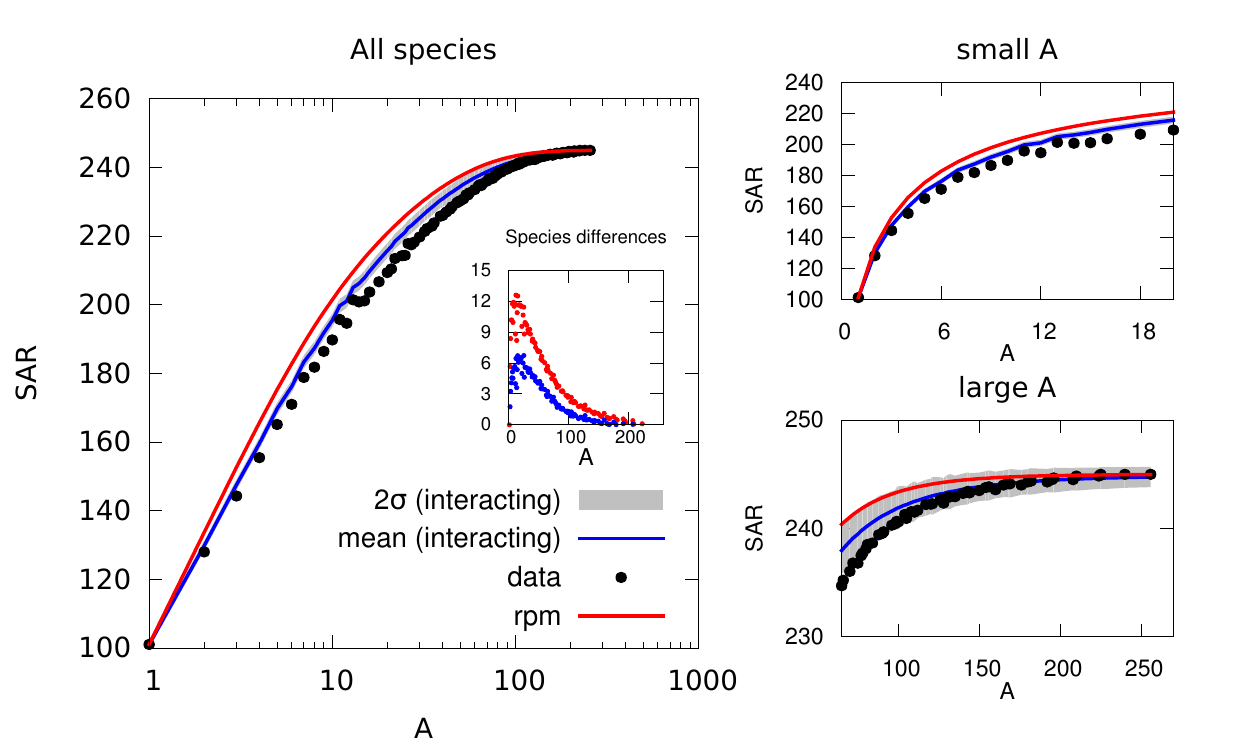} }
  \caption{\small{\textbf{SAR for the ecosystem}  The SAR over all the sampled areas (left) with the inset that shows the differences between data and the two models (red for random placement and blue for interacting model); the right column shows a magnification for small (top) and large regions (bottom). The grey area represents the $2\sigma$ confidence interval for the interacting model. The panels on the right represent respectively the magnifications at small and large areas.}}
  \label{fig:SAR}
\end{figure}

Figure~\ref{fig:SAR} shows the SAR for the BCI ecosystem. In both the random placement model as well as the interacting one, the SAR obviously converges to the total species richness due to the fact that the probability to find a species on the largest surveyed area is one. In order, to quantify the reliability of the predicted SAR for both the interacting and random models we evaluated the difference between the predicted species richness by the models and the one extracted from the data (inset Figure \ref{fig:SAR}). Although the differences between prediction and data in both models present a peak at an intermediate area, the interacting model consistently performs better. The discrepancy between the interacting model and the data in the intermediate spatial scale can be understood by the fact that the effective interaction J characterizing the clustering is obtained using the information at a spatial scale imposed by the nearest neighbor separation. To be more precise, species aggregate in different ways at different spatial scales \cite{Hartley2004,Plotkin2002} and this can have an effect on the effective space dependent couplings $J$.

\begin{figure}[h!]
  \centerline{ \includegraphics[width=1.2\linewidth,,height=1.2\linewidth,keepaspectratio]{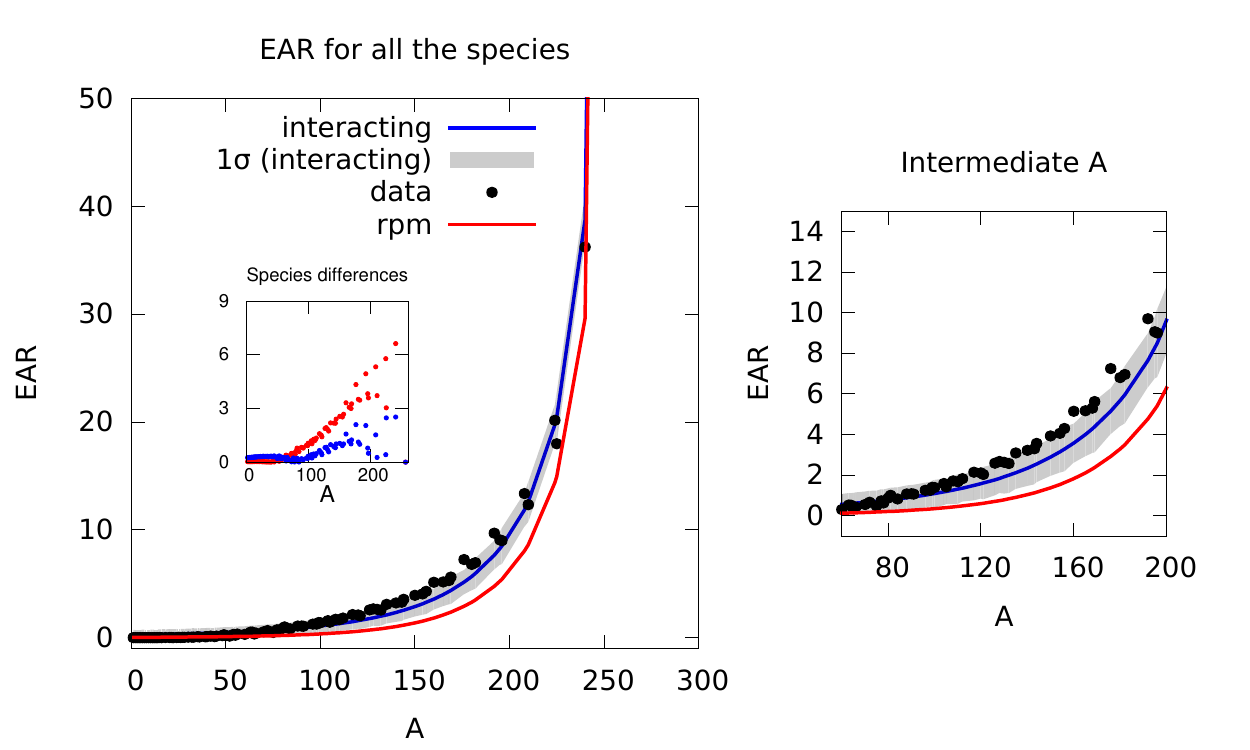} }
  \caption{\small{\textbf{EAR for the ecosystem.} The EAR over all the sampled areas. The interacting model (blue) and the random placement model (red) reach the same value for the largest $A$ because every species is completely contained in the surveyed area.}} 
  \label{fig:EAR}
\end{figure}

Figure \ref{fig:EAR} shows the EAR for the BCI ecosystem. The EAR is negligible at small areas, while it is ``forced'' to reach the total number of species in the largest plot (loosely speaking all the species are endemic in the largest area).
He and Hubbell~\cite{He2011}, using abundance data from rainforest ecosystems, showed a good agreement of the random placement EAR with data. However, we find that the interacting model significantly improves the predictions for the EAR with respect to the RPM. In particular, we find that the RPM model systematically underestimates the number of endemic species within a given area. 

\section{Conclusions}
The increased availability of presence/absence data of species occurrences has stimulated the formulation of models and methods to describe emergent spatial patterns. Most of the models that have been proposed using MaxEnt do not consider spatial features in an explicit manner. In order to understand and characterize spatial features in ecological communities, here we propose a spatially explicit maximum entropy model suitable for this kind of presence/absence data. Using only a few assumptions on the kind of information crucial to capture the spatial structure of the community, the model can extrapolate the SAR and EAR to larger spatial scales. The SAR for the interacting model presents discrepancies with the data in the range of intermediate sampled areas. This suggests that the effective interactions are scale dependent and call for an extension of the analysis to different spatial resolutions. In contrast, the EAR is fairly well reproduced by our MaxEnt spatial interaction model. Our approach is particularly suitable for studying rare species which are known to play a central role in ecosystem functioning and stability \cite{mouillot2013,suweis2013} - and predict the impact that habitat fragmentation may have on their conservation. Finally our model may be well suited for the analysis of communities at different spatial scales and the vital problem of upscaling.

\FloatBarrier

\appendix

\section{Relation between $h^\alpha$ and $\hat{M}^\alpha$ for the random placement model}
\label{app:hMrelation}
For the random placement model, imposing the condition that averaged quantities must reproduce the observed ones implies that:
\begin{equation}
  \langle M(\vec{\sigma}^\alpha)  \rangle_{h_{\alpha}} = N \frac{1}{1+e^{-h_\alpha}} = \hat{M}_\alpha .
\end{equation}
Defining $\hat{m}_\alpha=\frac{\hat{M}_\alpha}{N}$, the previous equation fixes $h_\alpha$ to be:
\begin{equation}
  \label{eq:h_M}
  h_\alpha = \ln \left( \frac{\hat{m}_\alpha}{1-\hat{m}_\alpha} \right).  
\end{equation}
Thus imposing only one constraint of reproducing the mean occurrence $\hat{M}^\alpha$ is equivalent to fixing the coupling $h_\alpha$.

\section{EAR for the case J = 0}
\label{app:EAR_rpm}
Starting from $ EAR_\alpha(a) = \Big \langle \mathlarger{\delta}_{\sum\limits_{i \in a^c} \sigma_i,0} \Big \rangle_{\boldsymbol{g_\alpha}} $, one can explicitly write down an expression for the EAR in Eq.~\eqref{eq:EAR_withZ}. In fact, using the definition of $ \langle \cdots \rangle_{\boldsymbol{g}_\alpha} $, we find:

\begin{equation}
  EAR_\alpha(a) = \frac{1}{Z(\boldsymbol{g_\alpha})} \sum \limits_{\boldsymbol{\sigma}} \mathlarger{\delta}_{\sum\limits_{i \in a^c} \sigma_i,0} e^{J_\alpha E(\boldsymbol{\sigma}) +  h_\alpha M(\boldsymbol{\sigma})} 
\end{equation}
The condition imposed by the Kronecker delta is satisfied only for the case $\sigma_i = 0$ for all the sites $i$ in $a^c$ and thus we obtain:
\begin{equation}
  EAR_\alpha(a) = \frac{1}{Z(\boldsymbol{g_\alpha})} \sum \limits_{\boldsymbol{\sigma} \in a} e^{J_\alpha E(\boldsymbol{\sigma}) +  h_\alpha M(\boldsymbol{\sigma})}
\end{equation}
When $J = 0$, each site is independent of the others, and so the expression above can be simplified to:

\begin{equation}
  EAR_\alpha(a) = \frac{1}{Z(h_\alpha)} \Big(\sum \limits_{\sigma} e^{h_\alpha \sigma}\Big)^{|a|}
\end{equation}
Taking into account Eq.~\ref{eq:h_M} and the fact that $ Z(h_\alpha) = Z_A(h_\alpha) $, the above expression gives us:
\begin{equation}
  EAR_{RPM}^\alpha(a) = \left( \frac{1}{1+e^{h_\alpha}} \right)^{A-|a|}
\end{equation}

\bibliographystyle{unsrt}
\bibliography{biblio_MaxentEcology}

\begin{thebibliography}{10}

\bibitem{condit2002}
Richard Condit, Nigel Pitman, Egbert~G Leigh, J{\'e}r{\^o}me Chave, John
  Terborgh, Robin~B Foster, Percy N{\'u}nez, Salom{\'o}n Aguilar, Renato
  Valencia, Gorky Villa, et~al.
\newblock Beta-diversity in tropical forest trees.
\newblock {\em Science}, 295(5555):666--669, 2002.

\bibitem{Volkov2005}
I~Volkov, JR~Banavar, F~He, S~Hubbell, and A~Maritan.
\newblock {Density dependence explains tree species abundance and diversity in
  tropical forests.}
\newblock {\em Nature}, 438(7068):658--61, December 2005.

\bibitem{Holts2006}
Robert~D Holt.
\newblock Emergent neutrality.
\newblock {\em Trends in ecology \& evolution}, 21(10):531--533, 2006.

\bibitem{Storch2012}
David Storch, Petr Keil, and Walter Jetz.
\newblock Universal species$\textendash$area and endemics$\textendash$area
  relationships at continental scales.
\newblock {\em Nature}, 488(7409):78--81, Jun 2012.

\bibitem{UNTB}
S~Hubbell.
\newblock {\em {The Unified Theory of Biodiversity and Biogeography}}.
\newblock Princeton Univeristy Press, 2001.

\bibitem{alonso2006merits}
David Alonso, Rampal~S Etienne, and Alan~J McKane.
\newblock The merits of neutral theory.
\newblock {\em Trends in Ecology \& Evolution}, 21(8):451--457, 2006.

\bibitem{blythe2007}
Richard~A Blythe and Alan~J McKane.
\newblock Stochastic models of evolution in genetics, ecology and linguistics.
\newblock {\em Journal of Statistical Mechanics: Theory and Experiment},
  2007(07):P07018, 2007.

\bibitem{Dewar2008}
Roderick~C Dewar and Annabel Port{\'e}.
\newblock Statistical mechanics unifies different ecological patterns.
\newblock {\em Journal of Theoretical Biology}, 251(3):389--403, 2008.

\bibitem{harte2011}
John Harte.
\newblock {\em Maximum entropy and ecology: a theory of abundance,
  distribution, and energetics}.
\newblock Oxford University Press, 2011.

\bibitem{rosindell2011}
James Rosindell, Stephen~P Hubbell, and Rampal~S Etienne.
\newblock < i> the unified neutral theory of biodiversity and biogeography</i>
  at age ten.
\newblock {\em Trends in ecology \& evolution}, 26(7):340--348, 2011.

\bibitem{suweis2012a}
S~Suweis, E~Bertuzzo, L~Mari, I~Rodriguez-Iturbe, A~Maritan, and A~Rinaldo.
\newblock On species persistence-time distributions.
\newblock {\em Journal of theoretical biology}, 303:15--24, 2012.

\bibitem{Chave2002}
J~Chave, H~C Muller-Landau, and S~A Levin.
\newblock {Comparing classical community models: Theoretical consequences for
  patterns of diversity}.
\newblock {\em American Naturalist}, 159(1):1--23, January 2002.

\bibitem{volkov2003}
Igor Volkov, Jayanth~R Banavar, Stephen~P Hubbell, and Amos Maritan.
\newblock {Neutral theory and relative species abundance in ecology.}
\newblock {\em Nature}, 424(6952):1035--7, August 2003.

\bibitem{alonso2004}
David Alonso and Alan~J McKane.
\newblock Sampling hubbell's neutral theory of biodiversity.
\newblock {\em Ecology Letters}, 7(10):901--910, 2004.

\bibitem{Azaele2006}
Sandro Azaele, Simone Pigolotti, Jayanth~R. Banavar, and Amos Maritan.
\newblock Dynamical evolution of ecosystems.
\newblock {\em Nature}, 444(7121):926--928, 2006.

\bibitem{volkov2007}
Hubbell S.~P. Volkov~I., Banavar Jayanth~R. and Maritan A.
\newblock Patterns of relative species abundance in rainforests and coral
  reefs.
\newblock {\em Nature}, 450(7166):45--49, November 2007.

\bibitem{Houchmandzadeh2008}
Bahram Houchmandzadeh.
\newblock Neutral clustering in a simple experimental ecological community.
\newblock {\em Physical review letters}, 101(7):078103, 2008.

\bibitem{bertuzzo2011}
Enrico Bertuzzo, Samir Suweis, Lorenzo Mari, Amos Maritan, Ignacio
  Rodr{\'\i}guez-Iturbe, and Andrea Rinaldo.
\newblock Spatial effects on species persistence and implications for
  biodiversity.
\newblock {\em Proceedings of the National Academy of Sciences},
  108(11):4346--4351, 2011.

\bibitem{suweis2012b}
Samir Suweis, Andrea Rinaldo, and Amos Maritan.
\newblock An exactly solvable coarse-grained model for species diversity.
\newblock {\em Journal of Statistical Mechanics: Theory and Experiment},
  2012(07):P07017, 2012.

\bibitem{Harte2003}
John Harte.
\newblock Ecology: Tail of death and resurrection.
\newblock {\em Nature}, 424(6952):1006--1007, 2003.

\bibitem{chayes1984inverse}
JT~Chayes, L~Chayes, and Elliott~H Lieb.
\newblock The inverse problem in classical statistical mechanics.
\newblock {\em Communications in Mathematical Physics}, 93(1):57--121, 1984.

\bibitem{He2011}
Fangliang He and Stephen~P Hubbell.
\newblock {Species-area relationships always overestimate extinction rates from
  habitat loss.}
\newblock {\em Nature}, 473(7347):368--71, May 2011.

\bibitem{legendre1989spatial}
Pierre Legendre and Marie~Jos{\'e}e Fortin.
\newblock Spatial pattern and ecological analysis.
\newblock {\em Vegetatio}, 80(2):107--138, 1989.

\bibitem{kerr2002local}
Benjamin Kerr, Margaret~A Riley, Marcus~W Feldman, and Brendan~JM Bohannan.
\newblock Local dispersal promotes biodiversity in a real-life game of
  rock--paper--scissors.
\newblock {\em Nature}, 418(6894):171--174, 2002.

\bibitem{Jaynes2003}
E.T. Jaynes and G.L. Bretthorst.
\newblock {\em Probability Theory: The Logic of Science}.
\newblock Cambridge University Press, 2003.

\bibitem{Burkoff2013}
N.~S. Burkoff, C.~Varnai, and D.~L. Wild.
\newblock Predicting protein -sheet contacts using a maximum entropy-based
  correlated mutation measure.
\newblock {\em Bioinformatics}, 29(5):580--587, Mar 2013.

\bibitem{Schneidman2006}
Elad Schneidman, Michael~J Berry, Ronen Segev, and William Bialek.
\newblock {Weak pairwise correlations imply strongly correlated network states
  in a neural population.}
\newblock {\em Nature}, 440(7087):1007--12, April 2006.

\bibitem{Harte2009a}
John Harte, Adam~B Smith, and David Storch.
\newblock {Biodiversity scales from plots to biomes with a universal
  species-area curve.}
\newblock {\em Ecology letters}, 12(8):789--97, August 2009.

\bibitem{Azaele2010}
S~Azaele, R~Muneepeerakul, A~Rinaldo, and I~Rodriguez-Iturbe.
\newblock Inferring plant ecosystem organization from species occurrences.
\newblock {\em Journal of theoretical biology}, 262(2):323--329, 2010.

\bibitem{Hoagland1997}
Bruce~W Hoagland and Scott~L Collins.
\newblock Gradient models, gradient analysis, and hierarchical structure in
  plant communities.
\newblock {\em Oikos}, pages 23--30, 1997.

\bibitem{Veech2006}
Joseph~A. Veech.
\newblock A probability-based analysis of temporal and spatial co-occurrence in
  grassland birds.
\newblock {\em J Biogeography}, 33(12):2145--2153, Dec 2006.

\bibitem{Volkov2009}
Igor Volkov, Jayanth~R Banavar, Stephen~P Hubbell, and Amos Maritan.
\newblock Inferring species interactions in tropical forests.
\newblock {\em Proceedings of the National Academy of Sciences},
  106(33):13854--13859, 2009.

\bibitem{banavar2010}
Jayanth~R Banavar, Amos Maritan, and Igor Volkov.
\newblock Applications of the principle of maximum entropy: from physics to
  ecology.
\newblock {\em Journal of Physics: Condensed Matter}, 22(6):063101, 2010.

\bibitem{Plotkin2000a}
J~B Plotkin, M~D Potts, N~Leslie, N~Manokaran, J~Lafrankie, and P~S Ashton.
\newblock {Species-area curves, spatial aggregation, and habitat specialization
  in tropical forests.}
\newblock {\em Journal of theoretical biology}, 207(1):81--99, November 2000.

\bibitem{morlon2008}
H{\'e}l{\`e}ne Morlon, George Chuyong, Richard Condit, Stephen Hubbell, David
  Kenfack, Duncan Thomas, Renato Valencia, and Jessica~L Green.
\newblock A general framework for the distance--decay of similarity in
  ecological communities.
\newblock {\em Ecology Letters}, 11(9):904--917, 2008.

\bibitem{Janzen1970}
Daniel~H Janzen.
\newblock Herbivores and the number of tree species in tropical forests.
\newblock {\em American naturalist}, pages 501--528, 1970.

\bibitem{Connell1971}
Joseph~H Connell.
\newblock On the role of natural enemies in preventing competitive exclusion in
  some marine animals and in rain forest trees.
\newblock {\em Dynamics of populations}, 298:312, 1971.

\bibitem{Coleman1981}
BD~Coleman.
\newblock {On random placement and species-area relations}.
\newblock {\em Mathematical Biosciences}, 215:191--215, 1981.

\bibitem{cover2012}
Thomas~M Cover and Joy~A Thomas.
\newblock {\em Elements of information theory}.
\newblock John Wiley \& Sons, 2012.

\bibitem{Hartley2004}
S.~Hartley, W.~E. Kunin, J.~J. Lennon, and M.~J.~O. Pocock.
\newblock Coherence and discontinuity in the scaling of specie's distribution
  patterns.
\newblock {\em Proceedings of the Royal Society B: Biological Sciences},
  271(1534):81--88, Jan 2004.

\bibitem{Plotkin2002}
Joshua~B Plotkin and Helene~C Muller-Landau.
\newblock Sampling the species composition of a landscape.
\newblock {\em Ecology}, 83(12):3344--3356, 2002.

\bibitem{mouillot2013}
David Mouillot, David~R Bellwood, Christopher Baraloto, Jerome Chave, Rene
  Galzin, Mireille Harmelin-Vivien, Michel Kulbicki, Sebastien Lavergne, Sandra
  Lavorel, Nicolas Mouquet, et~al.
\newblock Rare species support vulnerable functions in high-diversity
  ecosystems.
\newblock {\em PLoS biology}, 11(5):e1001569, 2013.

\bibitem{suweis2013}
Samir Suweis, Filippo Simini, Jayanth~R Banavar, and Amos Maritan.
\newblock Emergence of structural and dynamical properties of ecological
  mutualistic networks.
\newblock {\em Nature}, 500(7463):449--452, 2013.

\end{thebibliography}

\end{document}